\documentstyle[a4wide,12pt,epsf]{article}
\begin{document}
\begin{flushright}
\baselineskip=12pt
{SUSX-TH-98-003}\\
{hep-ph/yymmnn}\\
{March 1998}
\end{flushright}

\begin{center}
{\LARGE \bf SPARTICLE SPECTRUM AND DARK MATTER IN M-THEORY \\}
\vglue 0.35cm
{D.BAILIN$^{\clubsuit}$ \footnote
{D.Bailin@sussex.ac.uk}, G. V. KRANIOTIS$^{\spadesuit}$ \footnote
 {G.Kraniotis@rhbnc.ac.uk} and A. LOVE$^{\spadesuit}$ \\}
	{$\clubsuit$ \it  Centre for Theoretical Physics, \\}
{\it University of Sussex,\\}
{\it Brighton BN1 9QJ, U.K. \\}
{$\spadesuit$ \it  Department of Physics, \\}
{\it Royal Holloway and Bedford New College, \\}
{\it  University of London,Egham, \\}
{\it Surrey TW20-0EX, U.K. \\}
\baselineskip=12pt

\vglue 0.25cm
ABSTRACT
\end{center}

{\rightskip=3pc
\leftskip=3pc
\noindent
\baselineskip=20pt
The phenomenological implications of the eleven dimensional limit of 
$M$-theory (strongly coupled $E_8\times E_8$) are investigated. In particular 
we calculate the supersymmetric spectrum subject to constraints of correct 
electroweak symmetry breaking and the requirement that the lighest 
supersymmetric particle provides the dark matter of the universe. The 
$B$-soft term associated with the generation of a $\mu$ term in the 
superpotential is calculated and its phenomenology is discussed. }

\vfill\eject
\setcounter{page}{1}
\pagestyle{plain}
\baselineskip=14pt

In recent years it has become clear that the five perturbative 
string theories and the 11 dimensional supergravity are different 
limits in moduli space of a unique fundamental theory. This 
strongly indicates an  enormous degree of symmetry of the underlying 
theory and its intrincically non-perturbative nature. 
String duality correlates the six corners of the moduli space. 
The duality transformation involves Planck's constant $\hbar$ 
and is therefore  intrinsically quantum mechanical. 
One then might argue that before we have the complete picture of 
$M$(other)theory 
it is premature to make any attempt at phenomenology. 
However, it may be that the corners 
of the moduli space capture most of the features of the 
theory relevant for low-energy phenomenology \footnote{The 11-D limit 
of  $M$-theory is not gauge invariant so Horava and Witten have argued that 
quantum terms are needed to restore gauge invariance \cite{HORWIT}.
See however, M. Faux's 
argument for a consistent classical limit of $M$-theory \cite{FAUX}}.

One of the most interesting dualities 
(and the most relevant for low-energy phenomenology) 
is the one in which the low-energy limit of 
$M$-theory (i.e 11D-Supergravity) compactified on the line segment 
$I\sim S^1/\bf{Z}_2$ (i.e an orbifold) is equivalent to the 
strong coupling limit of the $E_8\times E_8$ heterotic string 
\cite{HORWIT}. 
In this picture on one end of the line segment 
of length $\pi \rho$ live the 
observable fields contained in 
the first $E_8$ while the hidden sector fields live in
the second $E_8$ factor
on the other end. Gravitational fields propagate in the bulk.

The main phenomenological virtue of such a framework is that due to 
the extra dimension one may obtain unification {\it of all} interactions 
at a scale $M_U\sim 3\times 10^{16} GeV$ consistent with experimental data 
for the low energy gauge couplings \cite{witten}. 
Also the analysis of gaugino condensation reveals that phenomenologically 
acceptable gaugino masses, 
comparable with the gravitino mass $m_{3/2}$, 
arise quite naturally in sharp contrast to  
the weakly coupled case where tiny gaugino masses were troublesome 
\cite{NILLES}.
It is therefore of great importance to investigate further the 
phenomenological implications of the 11-dimensional low energy limit 
of $M$-theory and to determine any deviations from the weakly coupled 
case.

Several 
papers have recently  analyzed  
the effective supergravity and the soft supersymmetry-breaking terms 
emerging in this framework
\cite{ANTO,NANO,NILLES,DUDAS,STEVE,
CHOI,LUKAS,LI}. Some properties of the sparticle spectrum 
which depend only on the boundary conditions and not on the details of 
the electroweak symmetry breaking, have been discussed in \cite{CHOI}.
However, a detailed analysis of the spectrum 
subject to the constraint of correct electroweak 
symmetry breaking 
\cite{Tam:Rad}, as well as  other phenomenological
implications, has not been performed. It is the purpose of this paper to 
investigate the phenomenological implications of $M$-theory relevant 
to accelerator experiments and the cosmological properties of the lightest 
supersymmetric particle (LSP) as well as the 
prospects for its detection in 
underground non-baryonic dark matter experiments.

The soft supersymmetry-breaking 
terms are determined by the following functions of the effective 
supergravity theory \cite{LUKAS,CHOI}:
\begin{eqnarray}
K&=&-ln(S+\bar{S})-3ln(T+\bar{T})+\Biggl(\frac{3}{T+\bar{T}}+
\frac{\alpha}{S+\bar{S}}\Biggr)|C|^2, \nonumber \\
f_{E_6}&=&S+\alpha T, \;\;f_{E_8}=S-\alpha T, \nonumber \\
W&=&d_{pqr}C^p C^q C^r
\label{mfunc}
\end{eqnarray}
where $K$ is the K$\rm{\ddot{a}}$hler potential, $W$ the 
perturbative superpotential, and 
$f_{E_6}, f_{E_8}$ are the gauge kinetic functions for the observable 
and hidden sector gauge groups $E_6$ and $E_8$ respectively. 
Also $S,T$ are the dilaton and 
Calabi-Yau moduli fields and $C^{\alpha}$ charged matter fields.
The superpotential
and the gauge kinetic functions are exact up to non-perturbative effects. 
The integer $\alpha=\frac{1}{8\pi^2}\int \omega \bigwedge 
[tr(F\bigwedge F)-\frac{1}{2}tr(A\bigwedge A)]$ for the 
K$\rm{\ddot a}$hler form 
$\omega$ is normalized as the generator of the integer (1,1) cohomology.

Given eqs(\ref{mfunc}) one can determine 
\cite{CHOI,LUKAS} the soft supersymmetry breaking terms
for the 
observable sector 
gaugino masses $M_{1/2}$, scalar masses $m_0$ and trilinear scalar 
masses $A$
as functions of the auxiliary fields $F_S$ and $F_T$ of the moduli 
$S,T$ fields respectively.
\footnote{We assume very small 
$CP$-violating phases in the soft terms. This assumption 
is supported by the CP-structure 
of the soft terms in  T-duality invariant 
string compactifications at least for the $T$-moduli sector \cite{us}.}
\begin{eqnarray}
M_{1/2}&=&
\frac{\sqrt{3}Cm_{3/2}}{(S+\bar{S}+\alpha(T+\bar{T})}
\Biggl((S+\bar{S})\sin\theta+\frac{\alpha(T+\bar{T})\cos\theta}
{\sqrt{3}}\Biggr), \nonumber \\
m_0^2&=&V_0+m_{3/2}^2-\frac{3 m_{3/2}^2 C^2}{3(S+\bar{S})+
\alpha(T+\bar{T})} \nonumber \\
&\times&\Biggl\{\alpha(T+\bar{T})\Biggl(2-
\frac{\alpha(T+\bar{T})}{3(S+\bar{S})+\alpha(T+
\bar{T})}\Biggr)\sin^2 \theta \nonumber \\
&+&(S+\bar{S})\Biggl(2-
\frac{3(S+\bar{S})}{3(S+\bar{S})+\alpha(T+
\bar{T})}\Biggr)\cos^2 \theta \nonumber \\
&-&\frac{2\sqrt{3}\alpha(T+\bar{T})(S+\bar{S})}
{3(S+\bar{S})+\alpha(T+\bar{T})}\sin\theta\cos\theta\Biggr\}, \nonumber \\
A&=&\sqrt{3}Cm_{3/2}\Biggl\{\Biggl(-1+
\frac{3\alpha(T+\bar{T})}{3(S+\bar{S})+\alpha(T+\bar{T})}\Biggr)
\sin\theta \nonumber \\
& &+\sqrt{3}\Biggl(-1+\frac{
3(S+\bar{S})}{3(S+\bar{S})+\alpha(T+
\bar{T})}\Biggr)\cos\theta\Biggr\}
\end{eqnarray}
while the  $B$-soft term associated with non-perturbatively generated 
$\mu$ term in the superpotential 
is given by:

\begin{eqnarray}
B_{\mu}&=&m_{3/2}\Bigl[-3 C \cos\theta -\sqrt{3}C\sin\theta  \nonumber \\
&+& \frac{6 C \cos\theta (S+\bar{S})}{3(S+\bar{S})+
\alpha (T+\bar{T})}  +\frac{2 \sqrt{3} C \sin\theta \alpha (T+\bar{T})}{
3(S+\bar{S})+\alpha (T+\bar{T})}-1 +
F^S \frac{\partial {ln \mu}}{\partial S}+F^T \frac{\partial {ln \mu}}
{\partial T}\Bigr] \nonumber \\
\label{beta}
\end{eqnarray}
In (2),(3) the auxiliary fields are parametrized as follows \cite{IBA:Spain}:
\begin{eqnarray}
F^S&=&\sqrt{3} m_{3/2} C (S+\bar{S}) \sin\theta, \nonumber \\
F^T&=&m_{3/2} C (T+\bar{T})\cos\theta
\end{eqnarray}
and $\theta$ is the goldstino angle which specifies the extent to which 
the supersymmetry breaking resides in the dilaton versus the moduli 
sector. Also $m_{3/2}$ is the gravitino mass 
and $C^2=1+\frac{V_0}{3 m_{3/2}^2}$ with $V_0$ the tree level vacuum 
energy density.
Note that in the limit $\alpha (T+\bar{T})\rightarrow 0$ we recover the soft
terms of the weakly coupled large $T$-limit of Calabi-Yau compactifications
\cite{IBA:Spain}.

We now consider the supersymmetric spectrum.
Our parameters are the goldstino angle $\theta$,
$\alpha(T+\bar{T})$,$sign \mu$ (which is not determined by the radiative 
electroweak symmetry breaking constraint), where 
$\mu$ is the Higgs mixing parameter in the 
low energy superpotential, 
and $\tan\beta $ (i.e the ratio of the two Higgs vacuum 
expectation values $\tan\beta=\frac{<H_2^0>}{<H_1^0>}$) if we leave  
$B$ a free parameter determined 
by  the minimization of the Higgs potential. 
If $B$ instead is given by
(\ref{beta}), one determines the value of $\tan\beta$. 
For this purpose  we take $\mu$ independent of $T$ and $S$
because of our lack of knowledge of 
$\mu$ in $M$-theory.
We also set $C=1$ 
in the above expressions assuming zero cosmological constant.
The soft masses start running from 
a mass $R_{11}^{-1}\sim 7.5 \times 10^{15} GeV$ with 
$R_{11}$ the extra $M$-theory dimension. This is perhaps the most 
natural choice. However values as low as $10^{13}$ GeV are possible 
and have been advocated by some authors \cite{NANO}. 
However, the recent analysis of \cite{NILLES} disfavours such 
scenarios. For the most part of our analysis we shall consider the 
former value of $R_{11}$, but we shall also comment
on the consequences of the latter.

Then using (2) (3) as boundary conditions for the soft terms,
one evolves the renormalization 
group equations down to the weak scale and determines 
 the sparticle 
spectrum compatible with the constraints of correct electroweak symmetry 
breaking and experimental 
constraints on the sparticle 
spectrum from unsuccessful 
searches at LEP,Tevatron etc, 
and also that the LSP provides a good dark matter candidate.

Electroweak symmetry breaking is characterized by the extrema equations 
\begin{eqnarray}
\frac{1}{2}M_Z^2&=&\frac{\bar{m}^2_{H_1}-\bar{m}^2_{H_2}\tan^2 \beta}
{\tan^2 \beta -1}-\mu^2 \nonumber \\
-B\mu&=&\frac{1}{2}(\bar{m}^2_{H_1}+\bar{m}^{2}_{H_2}+2\mu^2)\sin 2\beta
\end{eqnarray}
where 
\begin{equation}
\bar{m}^2_{H_1,H_2}\equiv m^2_{H_1,H_2}+\frac{\partial \Delta V}{
\partial {v^2_{1,2}}}
\end{equation}
and $\Delta V=(64 \pi^2)^{-1} {\rm {STr}} M^4[ln (M^2/Q^2)-\frac{3}{2}]$ 
is the 
one loop contribution to the Higgs effective potential. We include
contributions from the third generation of particles and sparticles.

Since $\mu^2 \gg M_Z^2$ for most of the allowed
region of the parameter space \cite{NATH}, the following 
approximate relationships 
hold at the 
electroweak scale for the 
masses of neutralinos and charginos, which of 
course depend on the details of 
electroweak symmetry breaking. 
\begin{eqnarray}
m_{\chi_1^{\pm}}\sim m_{\chi_2^0}\sim 2 m_{\chi_1^0} \nonumber \\
m_{\chi_{3,4}^{0}}\sim m_{\chi_2^{\pm}}\sim |\mu|
\label{neutralinos}
\end{eqnarray}
In (\ref{neutralinos}) $m_{\chi_{1,2}^{\pm}}$ are the chargino mass 
eigenstates and $m_{\chi_{i}^{0}},i=1\ldots 4$ are the four neutralino mass 
eigenstates with $i=1$ 
denoting the lightest 
neutralino. The former arise after diagonalization of the 
mass matrix. 
\begin{equation}
M_{ch}=\left(\begin{array}{cc}\\
M_2 & \sqrt{2} m_W \sin\beta \\
m_W \cos\beta & -\mu 
\end{array}\right)
\end{equation}

We consider first the extreme $M$-theory limit 
\footnote{ Note that in this limit the dilaton dominated scenario ,
i.e ($\theta=\frac{\pi}{2}$) is not feasible.} in which 
$\alpha(T+\bar{T})=2$.
An interesting case is when 
the scalar 
masses are much smaller than the gaugino masses at the unification scale.
\footnote{Note that the weakly coupled case scalar masses 
are comparable to gaugino mass $m_{0} \sim M$ at the unification scale.}
For instance this can be achieved by a 
goldstino angle $\theta=\frac{7\pi}{20}$.
In this case sleptons are much lighter 
compare to the gluino than in the weakly coupled 
Calabi Yau compactifications of the heterotic
string for any value 
of $\theta$. As a result for high values of $\tan\beta$ \footnote{In case 
we determine  $B$ from the electroweak 
symmetry breaking $\tan\beta$ is a free parameter;
also in this case  
$M_{\tilde{g}}:m_{\tilde{L}_L}:m_{\tilde{e}_R}\sim
1:0.25:0.14$} there is a possibility 
that right handed selectrons or the lightest stau 
mass eigenstate become the 
LSP. 
The stau mass matrix is given by the expression 

\begin{equation}
{\cal M}_{\tau}^2=\left(\begin{array}{cc} \\
{\cal M}_{11}^2 & m_{\tilde{\tau}}(A_{\tau}+\mu \tan\beta) \\
m_{\tilde{\tau}}(A_{\tau}+\mu \tan\beta) & {\cal M}_{22}^2
\end{array}\right)
\end{equation}
This of course is phenomenologically unacceptable and results in 
strong constraints in the sparticle spectrum. In fig.1 we plot the 
critical value of gravitino mass $m_{3/2}^{c}$ above (below)  which 
$m_{\tilde{\tau}_2}$ or $m_{\tilde{e}_R}<m_{\chi_1^0}$ 
$(m_{\tilde{\nu}}<43GeV)$ versus $\tan\beta$. The acceptable 
parameter space lies between the upper and lower bound 
in fig.1. 
One can see that 
because of the above contrainst 
$\tan\beta \leq 13$ in this M-theory limit because the acceptable 
parameter space vanishes for larger values of $\tan\beta$.
We also plot in fig.2 the critical chargino mass $m_{\chi_1^{\pm}}$ 
,above which the right handed seleptons become the LSP, versus 
$\tan\beta$. In the same figure we also draw the experimental lower 
bound
for the chargino mass from LEP of about $83$ GeV (horizontal line). 
The $\tan\beta$ dependence of these constraints may be understood from the 
$D$-term contribution to the $\tilde{e}_R$ and $\tilde{\nu}$ mass 
formulas
\begin{equation}
\tilde{m}^2_i=c_i m_{3/2}^{2}-d_i \frac{\tan^2 \beta-1}{\tan^2 \beta+1} 
M^2_W
\end{equation}
where the $c_i$ are some RGE-dependent constants and 
$d_{\tilde{e}_R}=-\tan^2 \theta_W<0$ whereas $d_{\tilde{\nu}}=
\frac{1}{2}(1+\tan^2 \theta_W)>0$.

In the allowed region of fig.1 the LSP is the lightest 
neutralino $\chi_{1}^{0}$.
Assuming $R$-parity conservation 
the LSP is stable  and consequently 
can provide a good dark matter candidate. 
It is a linear combination of the superpartners of the 
photon, $Z^0$ and neutral-Higgs bosons,
\begin{equation}
\chi_1^0=c_1 \tilde{B}+c_2\tilde{W}^3+c_3\tilde{H}_1^0+c_4\tilde{H}_2^0
\end{equation}
The neutralino $4\times 4$ mass matrix can be written as
$$\left(\begin{array}{cccc} \\
M_1 & 0 & -M_Z A_{11} & M_Z A_{21} \\
0  & M_2 & M_Z A_{12} &-M_Z A_{22} \\
-M_Z A_{11} & M_Z A_{12} & 0 & \mu \\
M_Z A_{21} & -M_Z A_{22} & \mu & 0
\end{array}\right)$$ 
with 
$$\left(\begin{array}{cc} \\
 A_{11} & A_{12} \\
A_{21} & A_{22}\end{array}\right)= 
\left(\begin{array}{cc} \\
\sin\theta_{W} \cos\beta & \cos \theta_W \cos\beta \\
\sin\theta_W \sin\beta & \cos\theta_W \sin\beta 
\end{array}\right)$$

In fact the lightest 
neutralino in this model is almost a pure  bino ($\tilde {B}$),
which means $f_g\equiv |c_1|^2+|c_2|^2\gg 0.5$.
Most cosmological 
models predict that the relic abundance of neutralinos \cite{ARNO} 
satisfies
\begin{equation}
0.1\leq \Omega_{LSP} h^2 \leq 0.4
\label{COSMO}
\end{equation}

We calculated the relic abundance of the lightest neutralino using standard 
technology \cite{MARK} and found strong constraints on 
the resulting spectrum. In particular  for $\mu<0$ the lower limit on the 
relic abundance  results in 
a lower limit on the gravitino mass of about 200 GeV. In figs(3-6). 
we plot the 
relic abundance of the lightest neutralino versus the gravitino 
mass for various values of 
$\tan\beta$ for goldstino angle of $\theta=\frac{7\pi}{20}$. We also note 
that for the allowed parameter space $\Omega_{LSP} h^2$ never 
exceeds the upper limit of 0.4. However, for other values of the goldstino 
angle the upper limit on $\Omega_{LSP} h^2$ 
can constrain the gravitino mass. 
The lower limit also constraints the allowed values of the $\tan\beta$ 
parameter even further. For instance one can see from the plots 
that $\tan\beta<10$ in order that $\Omega_{LSP} h^2 \geq 0.1$.
In the allowed physical region direct 
detection rates are of order $10^{-2}-10^{-4} events/Kg/day$. 
The lightest Higgs $m_h$ is in the range 
$87GeV\leq m_h \leq 115 GeV$, while the neutralino mass is in the 
range $77GeV \leq m_{\chi_0} \leq 195 GeV$. 
 At this point 
is worth    
 mentioning
that if one chooses to run 
the soft masses from the mass 
$R_{11}^{-1}\sim O(10^{13})GeV$ instead of 
$7.5 \times 10^{15} GeV$ the cosmological constraint (
\ref{COSMO}) is very powerful and 
eliminates all of the parameter space since the relic abundance is always 
much smaller than 0.1, when $\theta=\frac{7\pi}{20}$.
For $\mu>0$ the maximum gravitino mass 
above which $m_{{\tilde e}_R}<m_{\chi_1^0}$
 or $m_{{\tilde \tau}_2}<m_{\chi_1^0}$ is smaller for fixed $\tan\beta$.
Clearly this novel $M$-theory limit provide us with a phenomenology 
distinct from the weakly coupled case \footnote{Note however, that this 
M-theory limit is somewhat similar to  the O-I orbifold model 
\cite{IBA:Spain,BK} and for a particular 
goldstino angle , different from the dilaton-dominated limit. 
On the other hand the O-I model has non-universal soft supersymmetry-
breaking terms at the string scale.}  
which should be a subject of experimental scrutiny.

For other values of the goldstino angle 
for which scalar masses are comparable to the 
gaugino masses (a case which is more similar to 
the weakly coupled limit \cite{IBA:Spain} in which 
$m_0\sim \frac{1}{\sqrt{3}} M_{1/2}$) we do not obtained constraints from
the bounds on the mass of  
right handed selectrons  and staus, but in this case the upper 
limit on the relic 
abundance 
leads to an upper limit on  the gravitino mass. For instance, for a
goldstino angle of $\theta=\frac{\pi}{4}$ (see fig.7) and 
$\tan\beta=2.5$, the requirement that 
$\Omega_{LSP} h^2 \leq 0.4$ results in $m_{3/2}\leq 365$ GeV. This results in 
an upper limit of the lightest Higgs $m_h\leq 100 GeV$. The lower limit 
is now $m_{3/2}\geq 115$ GeV. However, the LEP limit on the 
chargino mass of 82 GeV requires that $m_{3/2}\geq 150$ GeV. In this 
case the lightest neutralino is in the range 
$52 GeV\leq m_{\chi_1^0} \leq 148 GeV for  \mu<0$. For $\mu>0$ we have 
$65 GeV \leq m_{\chi_1^0} \leq 153 GeV$. Detection rates of the LSP 
for $^{73} Ge$ detector are in the range 
$6 \times 10^{-2}-10^{-4}events/Kg/day$ for $\mu<0$ and $O(10^{-3})-O(10^{-5}$
events/Kg/day for $\mu>0$. For higher $\tan\beta$ values on can obtain 
higher detection rates.

In conclusion, we have analyzed the supersymmetric 
spectrum and the properties 
of the lightest neutralino (LSP) in the 11-dimensional limit of M-theory. 
The most striking result 
,in the case of small scalar masses compare with  gluino masses, is that one 
obtains a limit on $\tan\beta\leq 13$, since above that value the 
right handed selectron or the lightest stau is the LSP, which is 
phenomenologically unacceptable since the LSP should be electrically neutral.
Also the cosmological constraint on the relic abundance of the LSP results
in a lower limit on the gravitino mass $m_{3/2}\geq 200$GeV. 
This further constrains  $\tan\beta$; $\tan\beta<10$.
In this case 
the upper limit on the relic abundance is not relevant since 
$\Omega_{LSP} h^2 < 0.4$ for goldstino angle $\theta=\frac{7\pi}{20}$ 
and for all the allowed values of $\tan\beta$. Also the lower 
limit on the relic abundance excludes the case of 
$R_{11}^{-1}\sim O(10^{13})$ GeV.
The scenario with $B_{\mu}$ given by (\ref{beta}),
and $\mu$ independent of, $S$  and, $T$ is excluded since one has 
to go to non-perturbative Yukawa couplings in order to obtain a value of 
$\tan\beta$ consistent with $B_{\mu}$ as in  (\ref{beta}).
For other values of the goldstino angle 
(which resemble more the weakly coupled 
Calabi-Yau compactifications for which 
$m_{0}=\frac{1}{\sqrt{3}}M_{1/2}$) the upper bound on the relic 
abundance results in an upper bound on the gravitino mass. For 
a goldstino angle, $\theta=\frac{\pi}{4}$ 
and $\tan\beta=2.5$ then we find $m_{3/2}\leq 365$GeV. 
Direct detection rates of the lightest neutralino are in 
the range of $10^{-1}-10^{-4} events/Kg/day$. 
The $M$-theory limit yields interesting phenomenology 
which should be the target of  experimental investigation.

\section*{Acknowledgements}
This research is supported in part by PPARC.

\newpage

\begin{figure}
\epsfxsize=6in
\epsfysize=8.3in
\epsffile{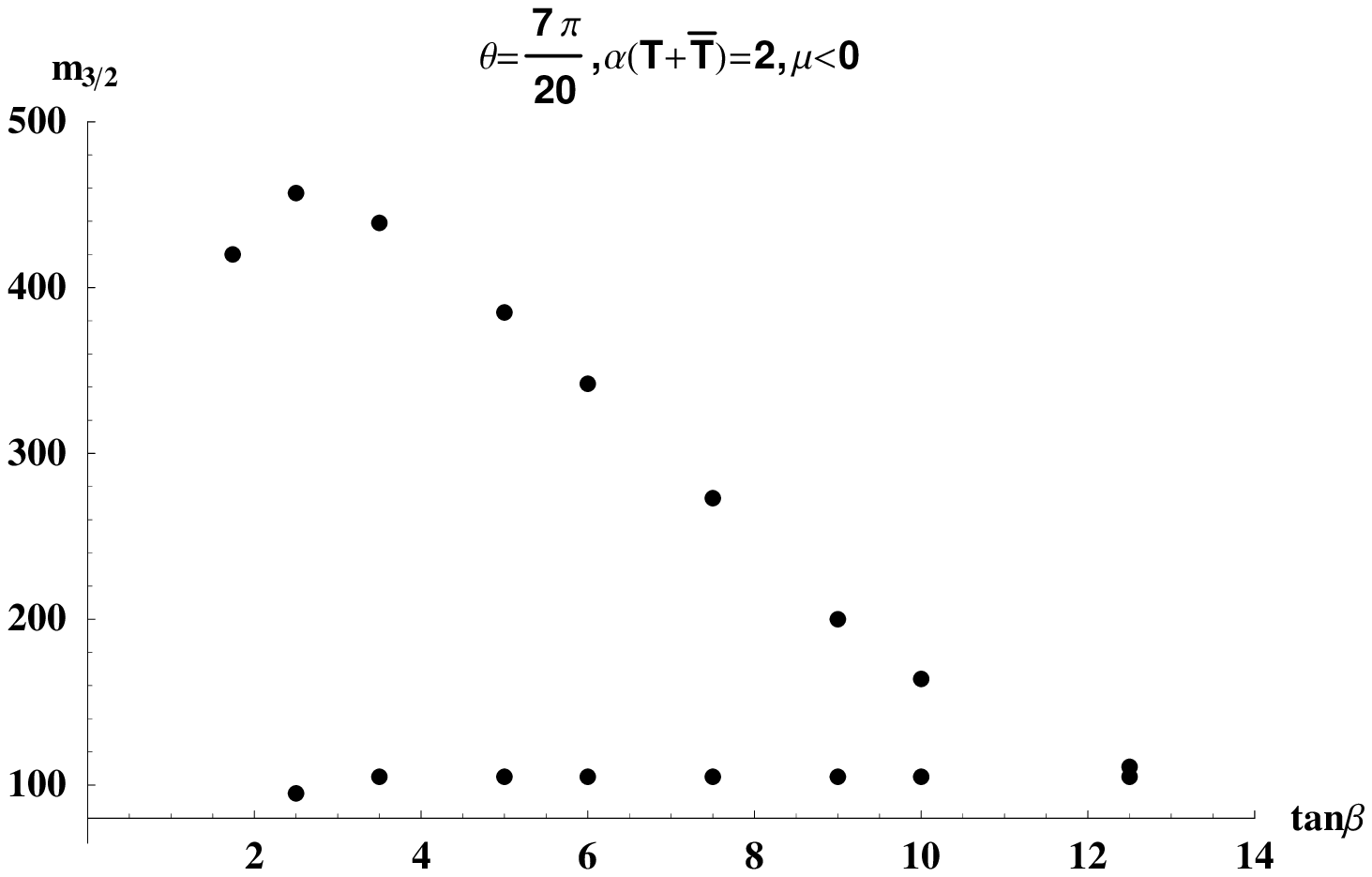}
\caption{Allowed parameter space in $m_{3/2},\tan\beta$ plane}
\end{figure}

\begin{figure}
\epsfxsize=6in
\epsfysize=8.3in
\epsffile{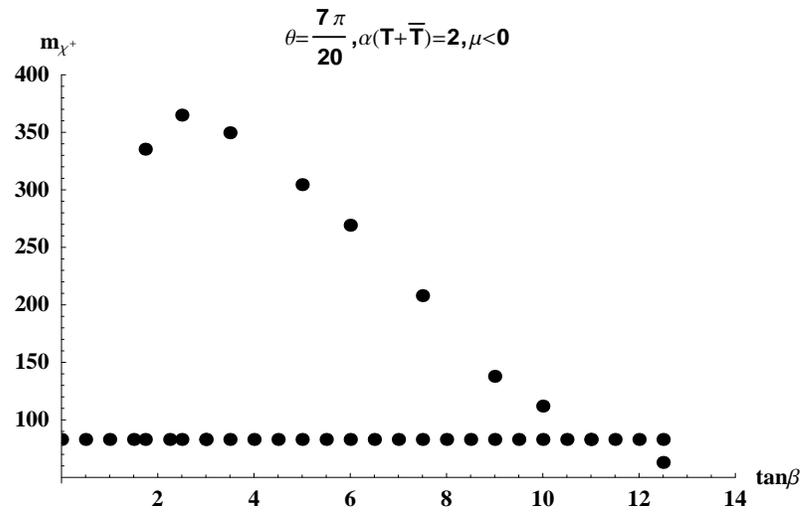}
\caption{Maximum lightest chargino mass vs $\tan\beta$}
\end{figure}

\newpage
\begin{figure}
\epsfxsize=6in
\epsfysize=8in
\epsffile{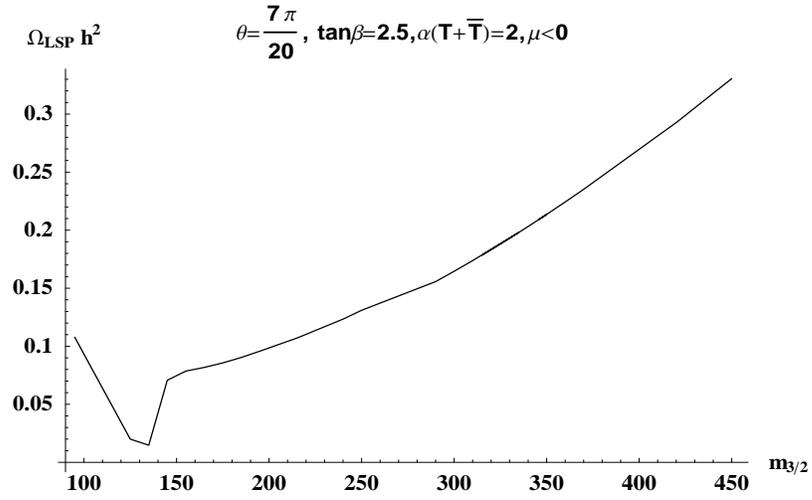}
\caption{Relic abundance versus $m_{3/2}$ for $\tan\beta=2.5,\mu<0$}
\end{figure}

\newpage
\begin{figure}
\epsfxsize=6in
\epsfysize=8.0in
\epsffile{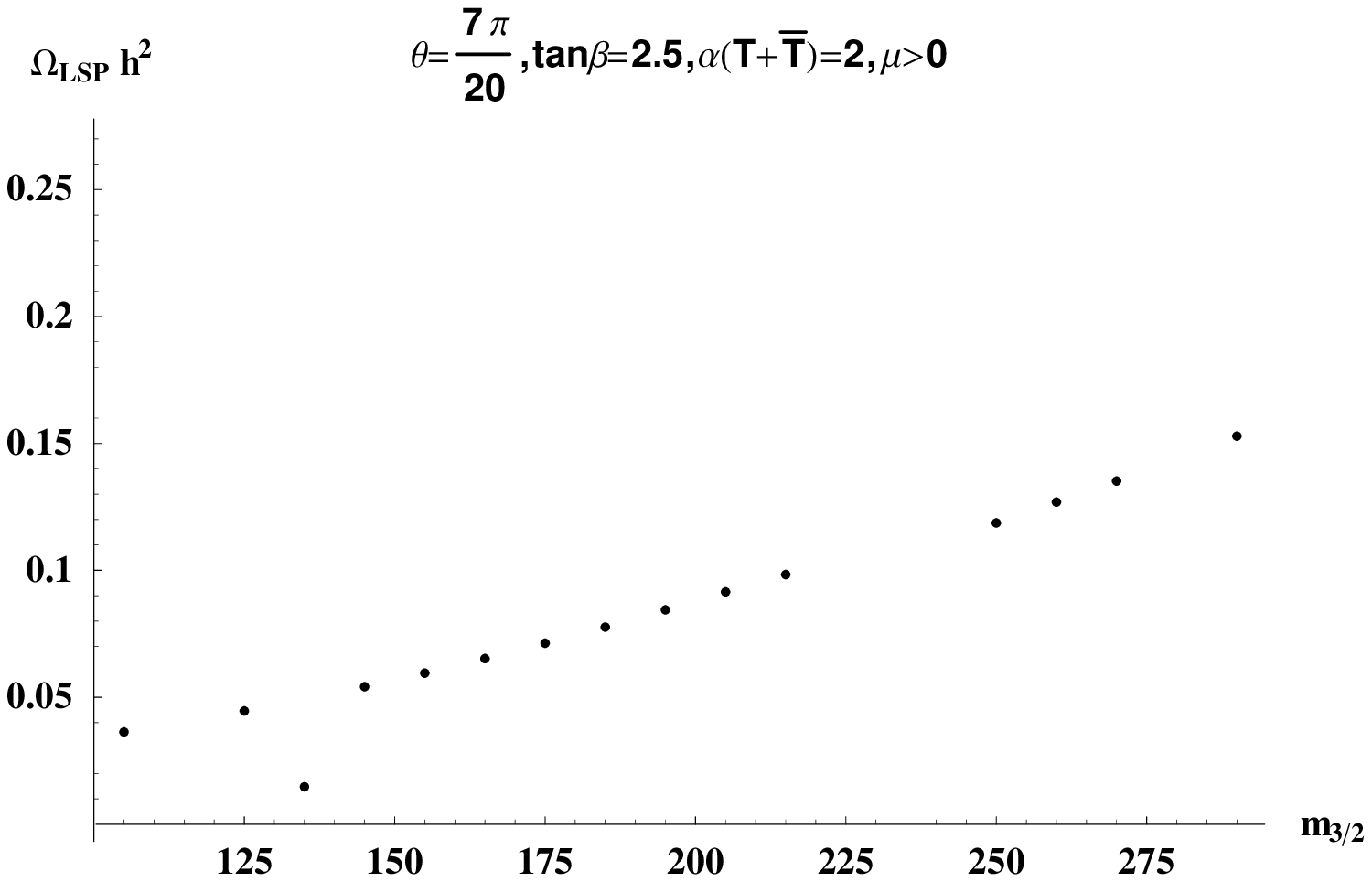}
\caption{Relic abundance of LSP vs $m_{3/2}$ for $\tan\beta=2.5,\mu>0$  }
\end{figure}

\newpage
\begin{figure}
\epsfxsize=6in
\epsfysize=8.5in
\epsffile{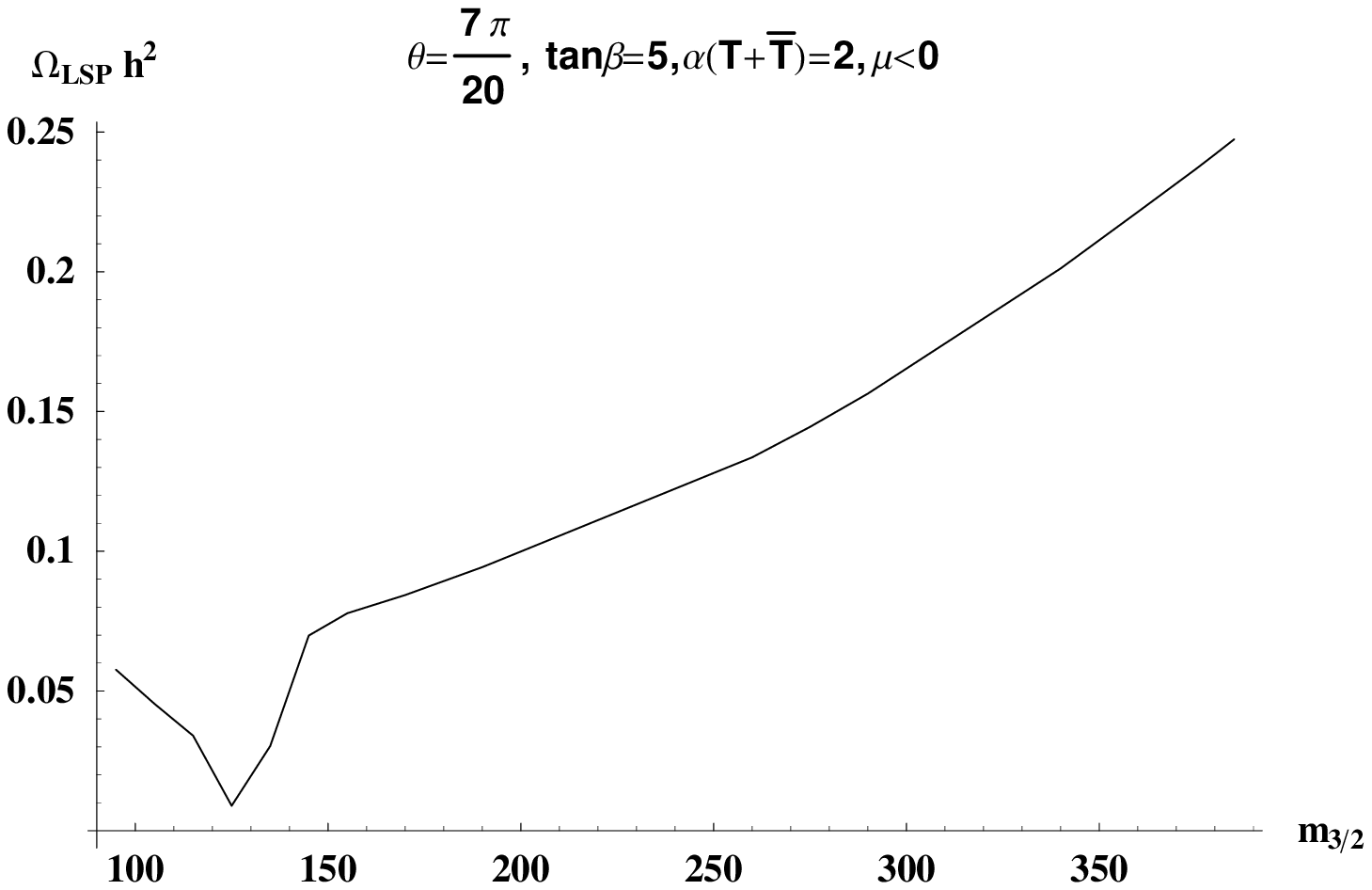}
\caption{Relic abundance of LSP vs $m_{3/2}$ for $\tan\beta=5, \mu<0$ }
\end{figure}


\newpage
\begin{figure}
\epsfxsize=6in
\epsfysize=8.5in
\epsffile{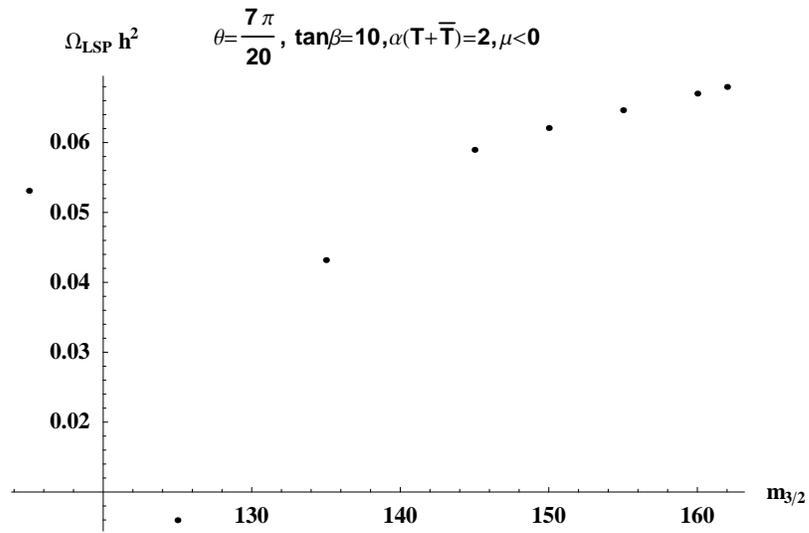}
\caption{Relic abundance of LSP in case of $\tan\beta=10$}
\end{figure}

\newpage
\begin{figure}
\epsfxsize=6in
\epsfysize=8.5in
\epsffile{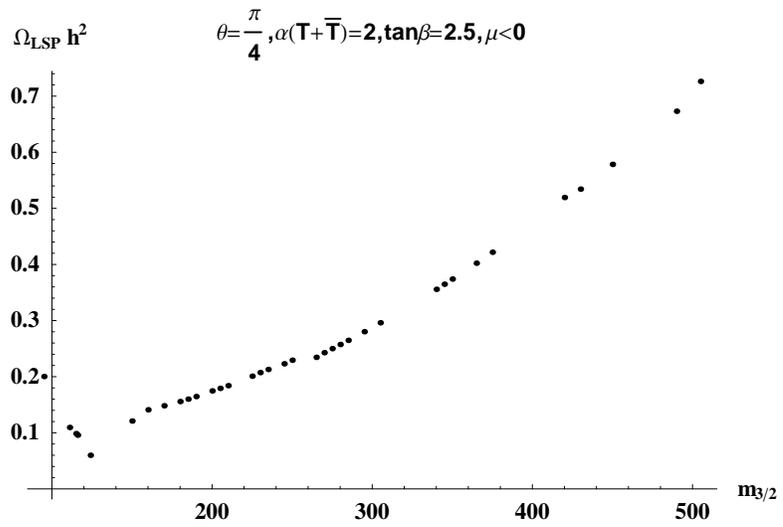}
\caption{Relic abundance of LSP in case of $\theta=\frac{\pi}{4}$ and 
$\tan\beta=2.5$}
\end{figure}

\end{document}